\documentclass[12p]{article}
\usepackage{hyperref}

\begin{document}

\input amssym.tex

\title{On the rest and flat limits of the scalar modes on the de Sitter spacetime}

\author{Ion I.  Cot\u{a}escu\thanks{E-mail:~~~cota@physics.uvt.ro}, 
Gabriel Pascu\thanks{E-mail:~~~gpascu@physics.uvt.ro} and Flavius Alin Dregoesc\thanks{E-mail:~~~dregoesc\_flavius@yahoo.com}\\
{\small \it West University of Timi\c{s}oara,}\\
{\small \it V.  P\^{a}rvan Ave.  4, RO-300223 Timi\c{s}oara, Romania}}

\maketitle

\begin{abstract}
The scalar mode functions on the spatially flat FLRW chart of the de Sitter spacetime are redefined in order to obtain a natural frequencies separation in the special local charts where the momentum vanishes, called here (natural) rest frames. This can be done since in such frames the mode functions are eigenfunctions of the energy operator. Moreover, the most general criteria for choosing the suitable phase factors of the mode functions able to simultaneously determine the correct rest and flat limits are established. These results are compared to the different original choices from the literature.
\end{abstract}

Pacs: 04.62.+v 


\newpage

\section{Introduction}


Inflationary cosmology has been in recent years a hot topic of debate and research, since it was proven that our Universe is expanding. The quantum field theory on an inflating background could give us information about how matter was created in the Universe, or even address the current dark matter enigma.

Quantum field theory is currently reliable only when formulated in an empty spacetime with zero curvature - the maximally symmetric Minkowski spacetime, the stage of special relativity. In case the cosmological constant does not vanish- the arena on which physical phenomena take place in an otherwise empty space, becomes the de Sitter spacetime. Since the de Sitter manifold is maximally symmetric, it has been viewed as a natural extension for special relativity for the case of non-vanishing cosmological constant. \cite{mignemi,aldrovandi2007sitter,aldrovandi2009physics} The free field theory on this manifold is relatively well known, expressed in several charts, and for different spins. \cite{nachtmann,borner_durr,chernikov_tagirov,cotaescu_crucean_pop} However, the opinions on how to build a correct interacting quantum field theory vastly differ. \cite{tagirov,kitamoto,polyakov,prokopec,cotaescu2010first} This ambiguity is bound to persist for a while, since measurement of the effects are impossible at the present time, because of the sensitivity required. \cite{aldrovandi2007sitter,cacciatori}

But the de Sitter quantities - at least the observable ones - must reduce, in the case when the cosmological constant vanishes to their Minkowksian counterparts, which can be verified by experiment. Also, the symmetry group reduces from the de Sitter group $SO(1,4)$ to the Poincar\'{e} group via the In\"{o}n\"{u}-Wigner contraction. Even the wavefunctions of matter fields, which are not observable quantities,  can be made to match, in some circumstances. 

The scalar mode functions in momentum representation, which are solutions of the Klein-Gordon equation on the spatially flat FLRW chart on de Sitter spacetime, are determined up to some phase factors, which can depend on momentum without dramatically affecting the physical meaning. Nevertheless, these play a crucial role in determining the rest frame limit (when the momentum vanishes), as well as the flat limit (the limit in which the cosmological constant parameter vanishes).

However, the addition of the cosmological parameter complicates things. Metric tensor components become time-dependent, the manifold is no longer asymptotically flat and even the notion of energy seems to be controversial since on de Sitter spacetime there is no global timelike Killing vector. Nevertheless, this is timelike in the light-cones of any chart including the spatially flat FLRW ones that are precisely the local charts which reduces to the pseudo-cartesian ones on the Minkowski spacetime. Therefore, the energy operator can be correctly defined \cite{nachtmann}, but this does not commute to the momentum components and, consequently, the energy-momentum dispersion does not hold. \cite{cotaescu_dirac1,cotaescu}

In special relativity, the rest frame of a particle represents the reference frame in which the particle has null momentum. In the quantum case, the rest frame is useful when considering processes like collisions or decays, because the wave functions in this limit have much simpler analytical expressions; one passes to the rest frame of a massive particle in order to simplify calculations. It is a tool which would benefit even on  curved manifolds where a momentum observable can be defined.

Fortunately, on the de Sitter manifold  the energy and  momentum can be defined as conserved quantities \cite{nachtmann,cotaescu_dirac1,cacciatori}, which allows the introduction of the rest frame in a classical approach. \cite{cacciatori} In the case of  quantum fields these are conserved observable of the $so(1,4)$ algebra generated by the de Sitter isometries. \cite{cotaescu} The quantum modes with non-vanishing momentum are not eigenfunctions of the energy operator because the dispersion relation between energy and momentum does not hold. However, in the rest frames, where the momentum vanishes,  the mode functions become eigenfunctions of the energy operator corresponding to positive or negative rest energies as defined in \cite{cotaescu}. Therefore the frequency separation can be done in the rest frame following the usual prescriptions as it was already done for the Dirac spinors. \cite{cotaescu_dirac,cotaescu2010first}

But how about the rest frame for a scalar quantum particle described by a wavefunction? In this paper the difficulties that stem from trying to define such a concept for the scalar case are explored, and a workaround is provided. The main objective of the paper is to study how the phase factors must be chosen in order to have correct rest and flat limits, discussing the most general conditions in which these limits can be performed and the peculiarities that are encountered. 

\section{Scalar fields on the de Sitter spacetime}

Let us consider the expanding part of de Sitter spacetime of Hubble constant $\omega$, endowed with the chart $\{t,{\bf x}\}$ of proper time and Cartesian space coordinates having the FLRW line element
\begin{equation}\label{mrw}
ds^2=g_{\mu\nu}(x)dx^{\mu}dx^{\nu}=dt^2-e^{2\omega t} (d{\bf x}\cdot d{\bf x})\,.
\end{equation}

The fields on the de Sitter manifold transform under the $SO(1,4)$ isometries according to covariant representations generated by conserved operators that form a covariant representation of the $so(1,4)$ algebra. Of special interest here are the momentum, ${\bf P}$, and energy, $H$, operators that read
\begin{equation}
{\bf P}=-i\nabla\,, \quad H= i\partial_t+\omega\, {\bf x}\cdot{\bf P}\,.
\end{equation}

These operators do not commute with each other and, therefore, cannot be put simultaneously in diagonal form. For this reason there are two different bases in which these are diagonal, namely the momentum and respectively energy representations. \cite{cotaescu_crucean_pop}

Using the aforementioned coordinates, consider the scalar field of mass $m$, whose time evolution is governed by the Klein-Gordon equation,
\begin{equation}\label{KG1}
\left( \partial_t^2-e^{-2\omega t}\Delta +3\omega
\partial_t+m^2\right)\phi(x)=0\,.
\end{equation}

The solutions with physical significance may be either square integrable functions or tempered distributions with respect to the scalar product
\begin{equation}\label{SP2}
\langle \phi,\phi'\rangle=i\int d^3x\, e^{3\omega t}\, \phi^*(x)
\stackrel{\leftrightarrow}{\partial_{t}} \phi'(x)\,.
\end{equation}

In the momentum representation (where the operators $P_i$ are diagonal) the scalar field can be expanded as
\begin{equation}\label{field1}
\phi(x)=\phi^{(+)}(x)+\phi^{(-)}(x)=\int d^3p \left[f_{\bf p}(x)a({\bf p})+f_{\bf p}^*(x)b^*({\bf p})\right] \,,
\end{equation}
where $a$ and $b$ are the wave functions of the momentum representation, while the quantum modes of the momentum basis
\begin{equation}
f_{\bf p}(x)=\frac{1}{2}\sqrt{\frac{\pi}{\omega}}e^{-i\chi({p})}\,e^{-3\omega
t/2}Z_k\left(\frac{p}{\omega}\,e^{-\omega t}\right) \frac{e^{i {\bf p}\cdot {\bf x}}}{(2\pi)^{3/2}}\,,
\end{equation}
are determined up to an arbitrary phases function $\chi({p})$ depending on $p=|{\bf p}|$. Here we denote
\begin{equation}\label{Z}
Z_k(s)=e^{-\pi k/2}H^{(1)}_{ik}(s)=e^{\pi k/2}H^{(1)}_{-ik}(s)\,,
\end{equation}
where $H^{(1)}$ are Hankel functions whose indices are given by
\begin{equation}\label{k}
k=\frac{M}{\omega}\,,\quad M=\sqrt{m^2-\lambda^2\omega^2}\,,
\end{equation}
provided $m>\lambda\omega$. The constant $\lambda$ depends on the type of coupling to gravity (for example $\lambda=\frac{3}{2}$ in minimal coupling or $\frac{1}{2}$ in the conformal one).

The fundamental solutions of positive frequencies, $f_{\bf p}$, and those of negative frequencies, $f_{\bf p}^*(x)$, satisfy the orthonormalization relations
\begin{eqnarray}
\langle  f_{\bf p},f_{{\bf p}'}\rangle=-\langle  f_{\bf p}^*,f_{{\bf
p}'}^*\rangle&=&\delta^3({\bf p}-{\bf p}')\,,\\
\langle f_{\bf p},f_{{\bf p}'}^*\rangle&=&0\,,
\end{eqnarray}
and the completeness condition
\begin{equation}\label{comp}
i\int d^3p\,  f^*_{\bf p}(t,{\bf x}) \stackrel{\leftrightarrow}{\partial_{t}}
f_{\bf p}(t,{\bf x}')=e^{-3\omega t}\delta^3({\bf x}-{\bf x}')\,,
\end{equation}
which do not depend on the choice of the phase function $\chi$.

On the contrary, the form of the Hamiltonian operator is strongly dependent on these phases. Indeed, we observe that the identity \cite{nachtmann,cotaescu_crucean_pop}
\begin{equation}\label{H}
(Hf_{\bf p})(x)=\left[-i\omega \left(p^i\partial_{p_i}+{\frac{3}{2}}\right)+\omega p^i\partial_{p^i}\chi(p)\right]f_{\bf
p}(x) \,,
\end{equation}
enables us to write,
\begin{eqnarray}
{\cal H}&=& \omega\int d^3 p\, {\cal E}(p)
\left[a^{\dagger}({\bf p}) a({\bf p})
+{b}^{\dagger}({\bf p}){b}({\bf p})\right]\nonumber\\
&+&\frac{i\omega}{2}\int d^3p\, p^i \left\{ \left[\, a^{\dagger}({\bf
p})\stackrel{\leftrightarrow}{\partial}_{p_i} a({\bf p})\right]+ \left[\,
b^{\dagger}({\bf p}) \stackrel{\leftrightarrow}{\partial}_{p_i} b({\bf
p})\right]\right\}\,,\label{Ham}
\end{eqnarray}
where we denote the 'apparent' energy as
\begin{equation}\label{Eap}
{\cal E}(p)=p^i\partial_{p^i}\chi(p)\,.
\end{equation}

\section{Rest and flat limits}

The limits of the mode functions in the rest frames (for $p\to 0$) can not be calculated using the limits of the Hankel functions (as is the case for a Dirac particle \cite{cotaescu_dirac}),
\begin{equation}
\lim_{{x}\to 0}x^{\nu} H^{(1)}_{\nu}(\alpha x)=-\lim_{{x}\to
0}x^{\nu} H^{(2)}_{\nu}( \alpha
x)=\frac{1}{i\pi}\left(\frac{2}{\alpha}\right)^{\nu}\Gamma(\nu)\,,
\end{equation}
since these  hold only for $\Re \nu >0$. The solution is to replace the functions
(\ref{Z}) with the new functions
\begin{equation}\label{Zpm}
Z^{(\pm)}_k(s)=e^{\mp\pi k/2}H^{(1)}_{0^+\pm ik}(s)=e^{\mp\pi k/2}\lim_{\epsilon\to 0} H^{(1)}_{\epsilon\pm ik}(s)\,,\quad \epsilon >0\,,
\end{equation}
obtaining thus the new mode functions $f_{\bf p}^{\pm}$ having now the phase functions $\chi^{\pm }$. In addition, we assume that these satisfy
\begin{equation}\label{I}
\lim_{p\to 0}\left[ \chi^{\pm}(p) \pm k \ln \left(\frac{p}{\omega}\right)\right] =\chi_0^{\pm}\,,
\end{equation}
where $\chi_0^{\pm}$ are arbitrary constants with respect to the variable $p$ (but can still depend on $\omega$).
Rigurously speaking, by introducing the $\epsilon$ variable, the problem becomes one to evaluate a limit of two variables. As such, one must be careful to specify on what curve the limit is taken, since generally limits taken along different curves need not coincide. By formally performing first the $p$ limit, and subsequently the $\epsilon$ one, it is understood that in terms of two variables, the limit is evaluated on a curve in the $(\epsilon,p)$ plane that identifies with the $\epsilon$ axis.

Under such circumstances, the limits for ${\bf p}\to 0$ do make sense and read
\begin{equation}
f_{\bf 0}^{\pm}=\lim_{{\bf p}\to 0}f_{\bf p}^{\pm}(x)=N_{\pm} e^{-iE^{\mp}_0 t}\,,\quad N_{\pm}=\frac{2^{\pm ik}}{2\sqrt{\pi\omega}}\frac{(-i)e^{-i\chi_0^{\pm}}}{(2\pi)^{3/2}}\Gamma(0^+\pm ik)e^{\mp\frac{k\pi}{2}}\,,
\end{equation}
where $E_0^{\pm}=\pm M-\frac{3}{2}i\omega$ are the particle $(+)$ and respectively antiparticle $(-)$ rest energies. \cite{cotaescu} \\

{ Hereby we understand that the correct choice is to consider $f^{-}_{\bf p}$ as the positive frequency mode function.}\\

Also, it is of note that the rest frame limit can be considered in this way only for fields with mass over the threshold $m>\lambda \omega$. Otherwise the pre-factor $\left( \frac{p}{\omega} \right)^\nu$ is not a phase factor vanishing in the rest frame.

It is instructive to show why the limit can't be evaluated on a curve along the $\epsilon=0$ axis. For pure imaginary orders, the Hankel function can be expanded at $z \sim 0$ as \cite{nist}
\begin{equation}
 H^{(1)}_{ik}(z) \sim -\frac{i}{\pi}\left[ \left(\frac{z}{2} \right)^{-ik}\Gamma(0^++ik) + \left(\frac{z}{2} \right)^{ik}\Gamma(0^+-ik) e^{\pi k} \right] \,,
\end{equation}
such that in our case we obtain for momenta of small moduli
\begin{equation}
f_{\bf p \sim 0}(x) \sim \left[ \left(\frac{p}{\omega} \right)^{ik} f_{\bf 0}^{-} +  \left(\frac{p}{\omega} \right)^{-ik} f_{\bf 0}^{+} \right] e^{i {\bf p}\cdot {\bf x}} \,,
\end{equation}
which can't have a limit fuction, in any way a phase would be chosen. This is the reason of the introduction of the parameter $\epsilon$, and why it is necessary in order to obtain a unambiguous limiting function.

The form of the phase function plays an important role in determining the flat limit of the mode functions. We study now how this must be chosen in order to recover the corresponding solution on the Minkowski spacetime
\begin{equation}\label{limf}
\lim_{\omega \to 0} f_{\bf p}^-(x)=\frac{1}{\sqrt{2E(p)}}\frac{e^{-iE(p)t+i {\bf p}\cdot {\bf x}}}{(2\pi)^{3/2}} \,,
\end{equation}
with $E(p)=\sqrt{m^2+p^2}$.  Moreover,  we can require the functions $a$ and $b$ to tend in the flat limit just to the corresponding functions of the Minkowski theory. This can be accomplished imposing the supplemental condition
\begin{equation}\label{II}
\lim_{\omega\to 0}{\cal E}^-(p)=\lim_{\omega\to 0}p\,\partial_p\,\chi^-(p)=E(p)\,.
\end{equation}
Then, in the flat limit, the first term of the operator (\ref{H}) becomes the usual energy operator, while the second one vanishes.

The function (\ref{Z}) can be approximated, using an asymptotic expansion \cite{bateman} as
\begin{equation}
Z^{(\pm)}_k \left( \frac{p}{\omega} e^{-\omega t} \right) \sim e^{-i\pi/4}\sqrt{\frac{2\omega}{\pi}}\left(M^2+p^2\,e^{-2\omega t}\right)^{-\frac{1}{4}} e^{i\phi(p,t)}\,,
\end{equation}
where
\begin{equation}
\phi(p,t)=
\frac{1}{\omega}\sqrt{M^2+p^2\,e^{-2\omega t}}-\frac{M}{\omega}\,{\rm arcsinh}\left(\frac{M\,e^{\omega t}}{p}\right)\,.
\end{equation}

Then the mode functions can be approximated as
\begin{equation}
f_{\bf p}^-(x)\sim e^{-i\pi/4}\frac{1}{\sqrt{2}}\left(M^2+p^2\,e^{-2\omega t}\right)^{-\frac{1}{4}} e^{i(\phi(p,t)-\chi^-(p))} e^{-3\omega t/2}\frac{e^{i {\bf p}\cdot {\bf x}}}{(2\pi)^{3/2}}\,.
\end{equation}

Now we observe that the phase function has a pole in $\omega=0$ allowing the series expansion
\begin{eqnarray}
\phi(p,t)&=&\frac{1}{\omega}\,E(p)-\frac{m}{\omega}\,{\rm arcsinh}\left(\frac{m}{p}\right)-t E(p)\nonumber\\
&+& \omega\left[\frac{p^2 t^2}{2E(p)}
+\frac{\lambda^2}{2m}{\rm arcsinh}\left(\frac{m}{p}\right)   \right] + {\cal O}(\omega^2)\,.
\end{eqnarray}
Therefore, the limit (\ref{limf}) holds only if the function $\chi^-$ satisfies
\begin{equation}\label{III}
\lim_{\omega\to 0}\left\{\chi^-(p)-\frac{1}{\omega}\left[\sqrt{m^2+p^2}-m\,{\rm arcsinh}\left(\frac{m}{p}\right)\right]\right\}=-\frac{\pi}{4}\,.
\end{equation}
The conclusion is that
{\em the phase functions must satisfy the conditions (\ref{I}), (\ref{II}) and
(\ref{III}).}

Obviously, there are many phase solutions. For example, a straightforward suitable choice is
\begin{equation} \label{choice}
\chi^-(p)=\frac{1}{\omega}\left[\sqrt{M^2+p^2}-M\,{\rm arcsinh}\left(\frac{M}{p}\right)\right]-\frac{\pi}{4}\,,
\end{equation}
when
\begin{equation}
{\cal E}^-(p)=\sqrt{M^2+p^2},
\end{equation}
and
\begin{equation}
\chi^-_0 = -\frac{\pi}{4}+k-k\ln(2k)\,.
\end{equation}

In order to check the consistency of the results, one should also verify the flat limit of the rest frame wavefunction $f_{\bf 0}^{\pm}$ leads to the correct result.
For that, expand the gamma function of large imaginary argument as a Stirling asymptotic expansion recovering the Minkowski wavefunction
\begin{equation}
\lim_{\omega \to 0}f_{\bf 0}^- =  \frac{1}{\sqrt{2m}} \frac{e^{-imt}}{(2\pi)^{3/2}}\,.
\end{equation}
It is of note that the similar limit for $f_{\bf 0}^+$ vanishes, since 
\begin{equation}
f_{\bf 0}^+ = e^{-\pi k} f_{\bf 0}^{- *}\,.
\end{equation}
This is another reason to use the $(-)$ solution in defining the rest frame instead of the $(+)$ one.

\section{Concluding remarks}

The conditions found for the limit of the vanishing cosmological constant in case of the de Sitter scalar wave functions agrees with the ones established when these plane wave solutions were first defined. \cite{nachtmann,borner_durr} The phase factor chosen then are of the form 
\begin{equation}
e^{-i\chi(p)}=e^{\frac{i\pi}{4}} e^{-\frac{i}{\omega}\sqrt{p^2+m_1(\omega)^2}} \left( \frac{p}{\sqrt{p^2+m_2(\omega)^2}-m_3(\omega)} \right)^{\frac{im_4(\omega)}{\omega}},
\end{equation}
with the explicit choices given in the following table. Note that the phases were considered for the $f_{\bf p}^{+}$ solution, since in the aforementioned references the $H^{(1)}_{ik}$ was taken into account as the preffered choice, in detriment of the one with negative pure imaginary order. 
\begin{table}[ht]
  \begin{center}
    \begin{tabular}{c|cccc}
         & $m_1$ & $m_2$ & $m_3$ & $m_4$ \\ \hline
				Nachtmann \cite{nachtmann} & $m$ & $m$ & $m$ & $M$ \\ 
				B\"{o}rner, D\"{u}rr \cite{borner_durr} & $m$ & $m$ & $M$ & $M$ \\ 
				relation (\ref{choice}) & $M$ & $M$ & $M$ & $M$ \\
    \end{tabular}
  \end{center}
\end{table}

This was a pretty general (but not the most general) way of choosing the phase function respecting condition (\ref{III}). 
As long as we are only concerned in taking the flat limit, any functions $m_i(\omega)$ satisfying $\lim \limits_{\omega \rightarrow 0} m_i(\omega)=m$, and additionally $\lim \limits_{\omega \rightarrow 0} \frac{dm_i(\omega)}{d\omega}=0$ are suitable. This goes to show the liberty of phase fixing.
However, when in addition we require the rest frame limit to exist, then $m_4(\omega)=M$ and $m_2(\omega)=m_3(\omega)$ (in the case of $(-)$ solution), or $m_2(\omega) \neq m_3(\omega)$ (in the case of $(+)$ solution). But in the latter $(+)$ case, as was discussed a subsequent flat limit from the de Sitter rest frame always vanishes.

Also, we have shown that there is a way of obtaining a rest frame result, using a small parameter $\epsilon$, introducing a way to overcome the difficulties arising in the definition for a rest frame wavefunction for a scalar field on the de Sitter spacetime. Summing up, the general conditions that the phases must satisfy in order for the wavefunctions to have correct limits have been established. While these factors are irrelevant in applications where the modulus squared of the wavefunction is used as a quantity, they do affect drastically the form of the Hamiltonian operator, and the existence of the aforementioned limits.


\begin{thebibliography}{0}

\bibitem{mignemi} Mignemi S., Doubly special relativity in de Sitter spacetime, {\it Annalen der Physik}, {\bf 522}(12): 924-940 (2010), \href{http://arxiv.org/abs/0802.1129}{arXiv:0802.1129 [gr-qc]}.

\bibitem{aldrovandi2007sitter} Aldrovandi R., Beltr\'{a}n Almeida J.P. and Pereira J.G., de Sitter special relativity, {\it Class. Quant. Grav., {\bf 24}(6): 1385-1404} (2007), \href{http://arxiv.org/abs/gr-qc/0606122}{arXiv:gr-qc/0606122}.

\bibitem{aldrovandi2009physics} Aldrovandi R. and Pereira J.G., Is Physics Asking for a New Kinematics?, {\it Int. J. Mod. Phys. D, {\bf 17}(13-14): 287-294} (2009), \href{http://arxiv.org/abs/0812.3438}{arXiv:0812.3438 [gr-qc]}.


\bibitem{nachtmann} Nachtmann O., Quantum theory in de-Sitter space, {\it Comm. Math. Phys.} {\bf 6}: 1-16 (1967).

\bibitem{borner_durr} B\"{o}rner G. and D\"{u}rr H.P., Classical and quantum fields in de Sitter space, {\it Il Nuovo Cimento A, Series 10} {\bf 64}(3): 669-714 (1969).

\bibitem{chernikov_tagirov} Chernikov N.A. and Tagirov E.A., Quantum theory of scalar field in the Sitter space-time, {\it Annales de l'institut Henri Poincar\'{e} (A) Physique th\'{e}orique} {\bf 9}(2): 109-141 (1968).

\bibitem{cotaescu_crucean_pop} Cot\u{a}escu I.I., Crucean C. and Pop A., The quantum theory of scalar fields on the de Sitter expanding universe, {\it Int. J. Mod. Phys. A} {\bf 23}(16-17): 2563-2577 (2008),  \href{http://arxiv.org/abs/0802.1972}{arXiv:0802.1972v1 [gr-qc]}.


\bibitem{tagirov} Tagirov E.A., Consequences of field quantization in de Sitter type cosmological models, {\it Annals of Physics} {\bf 76}(2): 561-579 (1973).

\bibitem{kitamoto} Kitamoto H. and Kitazawa Y., Boltzmann equation in de Sitter space, {\it Nucl. Phys. B} {\bf 839}(3): 552-579 (2010).

\bibitem{polyakov} Polyakov A. M., De Sitter Space and Eternity, {\it Nucl. Phys. B} {\bf 797}(1-2): 199-217 (2008).

\bibitem{prokopec} Prokopec T., Tsamis N.C. and Woodard R.P., Stochastic Inflationary Scalar Electrodynamics, {\it Annals Phys.} {\bf 323}(6): 1324-1360 (2008).

\bibitem{cotaescu2010first} Cot\u{a}escu I.I. and Crucean C., The de Sitter QED in Coulomb gauge: first order transition amplitudes (2010), \href{http://arxiv.org/abs/1007.4647}{arXiv:1007.4647 [gr-qc]}.


\bibitem{cacciatori} Cacciatori S., Gorini V., Kamenshchik A. and Moschella U., Conservation laws and scattering for de Sitter classical particles, {\it Class. Quantum Grav.} {\bf 25}(7): 075008 (2008), \href{http://arxiv.org/abs/0710.0315}{arXiv:0710.0315v2 [hep-th]}.


\bibitem{cotaescu_dirac1} Cot\u{a}escu, I.I., Polarized Dirac fermions in de Sitter spacetime, {\it Phys. Rev. D} {\bf 65}(8): 084008 (2002), \href{http://arxiv.org/abs/hep-th/0109199}{arXiv:hep-th/0109199}. 

\bibitem{cotaescu} Cot\u{a}escu I.I., The physical meaning of the de Sitter invariants, {\it GRG} {\bf 43}(6): 1639-1656 (2011),
\href{http://arxiv.org/abs/1006.1472}{arXiv:1006.1472v6 [gr-qc]}.

\bibitem{cotaescu_dirac} Cot\u{a}escu I.I., The free Dirac spinors of the spin basis on the de Sitter expanding universe, {\it Mod. Phys. Lett. A} {\bf 26}(22): 1613-1619 (2011), \href{http://arxiv.org/abs/1101.3164}{arXiv:1101.3164v4 [gr-qc]}.

\bibitem{nist} Olver F.W.J., Lozier D.W., Boisvert, R.F. and Clark C.W., {\it
NIST Handbook of Mathematical Functions}, Cambridge University Press, New York (2010).

\bibitem{bateman} Erdely A., Magnus W., Oberhettinger F. and Tricomi F.G., {\it Higher Transcendental Functions, volume II- based, in part, on notes left by Harry Bateman}, McGraw-Hill Book Co. (1953).



\end{thebibliography}
\end{document}